\RequirePackage{fix-cm}
\documentclass[twocolumn,epjc3]{svjour3}  
\RequirePackage{graphicx}
\usepackage{amsmath}
\usepackage{bm}
\journalname{Eur. Phys. J. E}

\begin{document}

\title{
Shear modulus and reversible particle trajectories of frictional granular materials under oscillatory shear
}


\author{Michio Otsuki\thanksref{e1,addr1}
        \and
	Hisao Hayakawa\thanksref{addr2} 
}

\thankstext{e1}{e-mail: otsuki@me.es.osaka-u.ac.jp}


\institute{Graduate School of Engineering Science, Osaka University, Toyonaka, Osaka 560-8531, Japan \label{addr1}
           \and
           Yukawa Institute for Theoretical Physics, Kyoto University, Sakyo-ku, Kyoto 606-8502, Japan \label{addr2}
}

\date{Received: date / Accepted: date}

\abstractdc{
  In this study, we numerically investigated the mechanical responses and trajectories of frictional granular particles under oscillatory shear in the reversible phase where particle trajectories form closed loops below the yielding point.
When the friction coefficient is small, the storage modulus exhibits softening, and the loss modulus remains finite in the quasi-static limit.
As the friction coefficient increases, the softening and residual loss modulus are suppressed. 
The storage and loss moduli satisfy scaling laws if they are plotted as functions of the areas of the loop trajectories divided by the strain amplitude and diameter of grains, at least for small values of the areas.
}

\maketitle

\section{Introduction}
\label{Intro}

Dense disordered materials, such as granular materials, foams, emulsions, and colloidal suspensions, behave like solids when the packing fraction $\phi$ exceeds the jamming point \cite{Hecke,Behringer}.
Under a small shear strain, the shear stress is proportional to shear strain, which is characterized by the shear modulus depending on $\phi$ \cite{OHern02,Tighe11,Otsuki17}. However, as the shear strain increases, the stress-strain relation becomes nonlinear \cite{Coulais,Otsuki14}.

The nonlinear stress-strain relation is believed to result from the yielding transition associated with plastic deformations \cite{Nagamanasa,Knowlton,Kawasaki16,Leishangthem,Clark,Boschan19}.
However, recent studies have shown that the mechanical response becomes nonlinear even when the system is in the reversible phase in which particle trajectories form closed loops below the yielding point \cite{Boschan,Nakayama,Kawasaki20,Bohy}.
Such a response is called (reversible) softening, where the storage modulus under oscillatory shear decreases as the strain amplitude increases.
In a previous paper \cite{Otsuki21}, we demonstrated that the loss modulus remains finite in the quasi-static limit when reversible softening occurs. We have clarified that the reversible softening and residual loss modulus originate from the loop trajectories of particles \cite{Lundberg,Schreck,Keim13,Keim14,Regev13,Regev15,Priezjev,Lavrentovich,Nagasawa,Das}.

Most previous numerical studies assumed frictionless particles, although realistic disordered materials consist of frictional grains. 
The friction causes drastic changes in rheology. 
For example, frictional particles exhibit discontinuous shear thickening \cite{Otsuki11,Chialvo,Brown,Seto,Fernandez,Heussinger,Bandi,Ciamarra,Mari,Grob,Kawasaki14,Wyart14,Grob16,Peters,Fall,Sarkar,Singh,Kawasaki18,Thomas} and shear jamming \cite{Bi11,Zhang08,Zhang10,Wang18,Zhao,Sarkar13,Sarkar16,Seto19,Pradipto,Otsuki20}, which hardly occurs in frictionless particles. 
Thus, it is natural to expect that the friction between the grains affects the mechanical responses and particle trajectories.

In this study, we numerically investigated the shear modulus of frictional granular materials under oscillatory shear.
In Sect. \ref{Setup}, we explain our model and setup.
In Sect. \ref{Reversible}, we present our numerical results of a single-cycle displacement and mean square displacement to distinguish the irreversible phase from the reversible phase of particle trajectories. 
We show how particle trajectories depend on the friction coefficient between the grains in Sect. \ref{Loop}.
In Sect. \ref{Response}, we illustrate the existence of scaling laws of the storage and loss moduli, at least, for small areas of reversible particles trajectories.
In Sect. \ref{Discussion}, we conclude and discuss our results.

\section{Model and setup}
\label{Setup}

Let us consider two-dimensional frictional granular particles with identical densities confined in a square box under oscillatory shear.
Particle $i$ with a diameter $d_i$ is driven by the SLLOD equation under the Lees--Edwards boundary condition \cite{Evans}:
\begin{eqnarray}
  \frac{d}{dt} {\boldsymbol r}_i & = & \dot \gamma(t) y_i \boldsymbol e_x + \frac{\boldsymbol p_i}{m_i}, \\
   \frac{d}{dt} {\boldsymbol p}_i & = & - \dot \gamma(t) p_{i,y} \boldsymbol e_x + \boldsymbol F_i,
\end{eqnarray}
where ${\boldsymbol r}_i=(x_i, y_j)$ and $\bm{p}_i=m_i(\dot{\bm{r}}_i-\dot\gamma(t)y_i) \bm{e}_x$ are the position and peculiar momentum of particle $i$ with mass $m_i$, shear rate $\dot \gamma(t)$, and the unit vector $\bm{e}_x$ along the $x$-direction, respectively.
The force $\boldsymbol F_i$ is given by:
\begin{equation}
  \boldsymbol F_i = \sum_{j \neq i} \left ( \boldsymbol F_{ij}^{\rm (n)} + \boldsymbol F_{ij}^{\rm (t)} \right ) \Theta(d_{ij} - r_{ij}),
\end{equation}
where $\boldsymbol F_{ij}^{\rm (n)}$ and $\boldsymbol F_{ij}^{\rm (t)}$ are
the normal and tangential forces between particles $i$ and $j$,
$d_{ij} = (d_i + d_j)/2$ is the average diameter,
and $r_{ij} = |\boldsymbol r_{ij}|$ is the distance between particles $i$ and $j$, with $\boldsymbol r_{ij} = \boldsymbol r_i - \boldsymbol r_j = (x_{ij},y_{ij})$.
Here, $\Theta(x)$ is the Heviside step function satisfying $\Theta(x)=1$ for $x>0$, and $\Theta(x) = 0$ otherwise.
To avoid crystallization, we constructed a dispersed system with an equal number of grains of two diameters ($d_0$ and $d_0/1.4$).

Then, we adopt the following model for the normal force:
\begin{equation}
 \boldsymbol F_{ij}^{\rm (n)} = -\left ( k^{\rm (n)} u^{\rm (n)}_{ij}
 + \eta^{\rm (n)} v^{\rm (n)}_{ij}\right ) \boldsymbol n_{ij}
\end{equation}
with $k^{\rm (n)}$ as the normal elastic constant, $\eta^{\rm (n)}$ as the normal viscous constant, and the normal unit vector is $\boldsymbol n_{ij} = \boldsymbol r_{ij}/r_{ij}$.
The normal relative displacement and velocity are, given by the following, respectively:
\begin{eqnarray}
u^{\rm (n)}_{ij} = r_{ij} - d_{ij},
\end{eqnarray}
and
\begin{eqnarray}
  v^{\rm (n)}_{ij} = \frac{d}{dt}u^{\rm (n)}_{ij} =
  \left ( \frac{d}{dt} \boldsymbol r_i - \frac{d}{dt} \boldsymbol r_j \right ) \cdot \frac{\boldsymbol r_{ij}}{r_{ij}}.
\end{eqnarray}
We adopt the following model for the tangential force
\begin{equation}
  \boldsymbol F_{ij}^{\rm (t)} = {\rm min} \left ( |\tilde F_{ij}^{\rm (t)}|, \mu F_{ij}^{\rm (n,el)} \right ) {\rm sgn} (\tilde F_{ij}^{\rm (t)}) \boldsymbol t_{ij},
\end{equation}
where $\boldsymbol t_{ij} = (-y_{ij}/r_{ij}, x_{ij}/r_{ij})$ is the tangential unit vector, and $\mu$ is the friction coefficient. Here, ${\rm min}(a,b)$ selects the smaller one between $a$ and $b$; ${\rm sgn}(x) = 1$ for $x \ge 0$ and ${\rm sgn}(x) = -1$ for $x < 0$. Furthermore, $F_{ij}^{\rm (n,el)} = - k^{\rm (n)} u^{\rm (n)}_{ij}$ is the elastic part of the normal force.
$\tilde F_{ij}^{\rm (t)}$ is given by
\begin{equation}
  \tilde F_{ij}^{\rm (t)} = - \left ( k^{\rm (t)} u_{ij}^{\rm (t)} + \eta^{\rm (t)} v_{ij}^{\rm (t)} \right )
\end{equation}
with $k^{\rm (t)}$ as the tangential elastic constant and $\eta^{\rm (t)}$ as the tangential viscous constant.
The tangential velocity $v_{ij}^{\rm (t)}$ is expressed as
\begin{equation}
  v_{ij}^{\rm (t)} = (\boldsymbol v_i - \boldsymbol v_j)\cdot \boldsymbol t_{ij} - (d_i \omega_i + d_j \omega_j)/2
\end{equation}
with $\omega_i$ as the angular velocity of particle $i$.
The tangential displacement $u_{ij}^{\rm (t)}$ satisfies $\dot u_{ij}^{\rm (t)} = v_{ij}^{\rm (t)}$ for $|\tilde F_{ij}^{\rm (t)}| <\mu F_{ij}^{\rm (n,el)}$, whereas $u_{ij}^{\rm (t)}$ remains unchanged for $|\tilde F_{ij}^{\rm (t)}| \ge \mu F_{ij}^{\rm (n,el)}$. We note that $u_{ij}^{\rm (t)}$ is set to zero if particles $i$ and $j$ are detached.
The time evolution of $\omega_i$ is given by
\begin{equation}
   I_i \frac{d}{dt} \omega_i = T_i 
\end{equation}
with the moment of inertia $I_i = m_i d_i^2 /8$, and torque $T_i = 
  - \sum_j \frac{d_i}{2} \boldsymbol F_{ij}^{\rm (t)} \cdot \boldsymbol t_{ij}$.

The particles were randomly placed with an initial packing fraction of $\phi_{\rm I} = 0.75$.
The system was slowly compressed until the packing fraction reached $\phi = 0.870$, which was sufficiently above the jamming point.
In each step of the compression, the packing fraction is increased by $\Delta \phi = 1.0 \times 10^{-4}$ with the affine transformation. Thereafter, the particles were relaxed to a mechanical equilibrium state with the kinetic temperature $T_{\rm K} = \sum_i p_i^2 /(mN) < T_{\rm th}$. 
Here, we chose $T_{\rm th} = 1.0 \times 10^{-8}k^{\rm (n)}d_0^2$.

After the compression, we apply the shear strain:
\begin{equation}
  \gamma(t) = \gamma_0 \sin \omega t
\end{equation}
at constant volume 
with a strain amplitude $\gamma_0$ and angular frequency $\omega$ for $N_{\rm c}$ cycles.
The shear rate is given by
\begin{equation}
  \dot \gamma(t) = \gamma_0 \omega \cos \omega t
\end{equation}
In the last cycle, we measured the storage and loss moduli, defined by \cite{Doi}
\begin{eqnarray}
  G' & = & \frac{\omega}{\pi} \int_0^{2\pi/\omega} dt \ \sigma(t) \sin \omega t / \gamma_0, \\
  G'' & = & \frac{\omega}{\pi} \int_0^{2\pi/\omega} dt \ \sigma(t) \cos \omega t / \gamma_0
\end{eqnarray}
with the (symmetric contact) shear stress as
\begin{equation}
  \sigma = - \frac{1}{2 L^2} \sum _i \sum_{j>i} (x_{ij} F_{ij,y}+y_{ij} F_{ij,x}).
\end{equation}
Here, we ignore the kinetic and asymmetric parts of the shear stress because they are less than $1 \%$ of $\sigma$.

The number of particles is $N=1000$, $k^{\rm (t)} = 0.2k^{\rm (n)}$, and $\eta^{\rm (n)} =\eta^{\rm (t)} = k^{\rm (n)}t_0$ with $t_0 = \sqrt{m/k^{\rm (n)}}$, where $m$ is the mass of a grain with diameter $d_0$.
This model corresponds to the restitution coefficient $e = 0.043$.
We adopt the leapfrog algorithm with the time step $\Delta t = 0.05 t_0$.
We chose $ \omega = 1.0 \times 10^{-4} t_0^{-1}$ as the quasi-static shear deformation because $G'$ and $G''$ do not depend on $ \omega$ for $\omega \le 1.0 \times 10^{-4} t_0^{-1}$.

\section{Single-cycle particle displacement and mean square displacement}
\label{Reversible}

\begin{figure}[htbp]
  \includegraphics[width=0.5\textwidth]{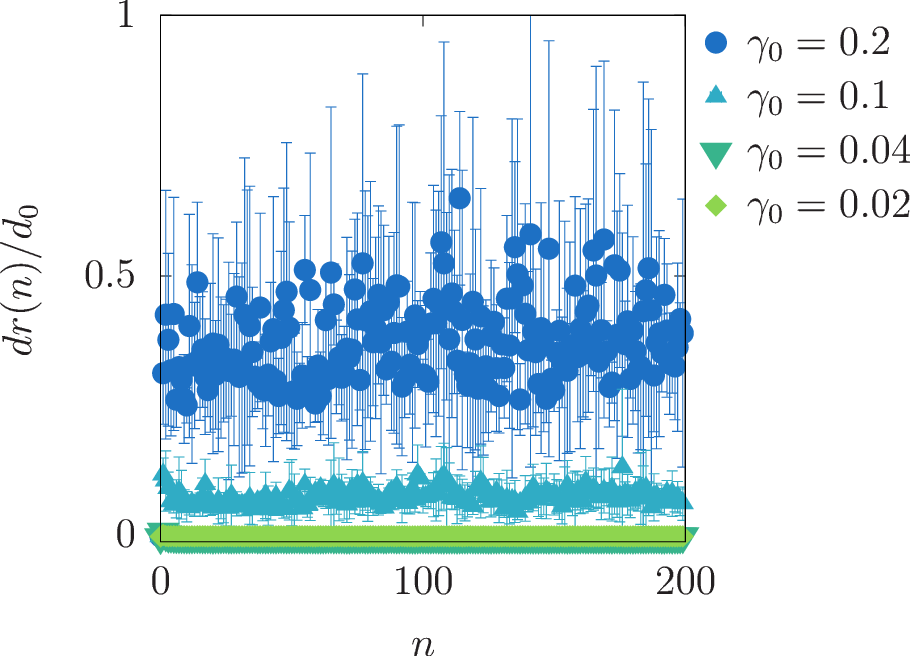}
\caption{Plots of the single-cycle displacement $dr(n)$ versus $n$ for various values of $\gamma_0$ with $\mu=1.0$.}
\label{dr}
\end{figure}

First, we introduce the single-cycle particle displacement as
\begin{equation}
  dr(n) = \left \langle \sum_{i=1}^N |\boldsymbol r_i(nT) - \boldsymbol r_i((n-1)T)| \right \rangle /N
\end{equation}
with the period $T = 2\pi/\omega$ and the ensemble average $\langle \cdot \rangle$.
We plot $dr(n)$ against $n$ for various values of $\gamma_0$ with $\mu=1.0$ in Fig. \ref{dr}.
For $\gamma_0 = 0.2$ and $0.1$, $dr(n)$ is finite and almost independent of $n$. 
The finite $dr(n)$ indicates that plastic deformations exist during the cycles.
For $\gamma_0 = 0.04$ and $0.02$, a negligibly small $dr(n)$ can be regarded as the reversible motion of particles.

\begin{figure}[htbp]
  \includegraphics[width=0.5\textwidth]{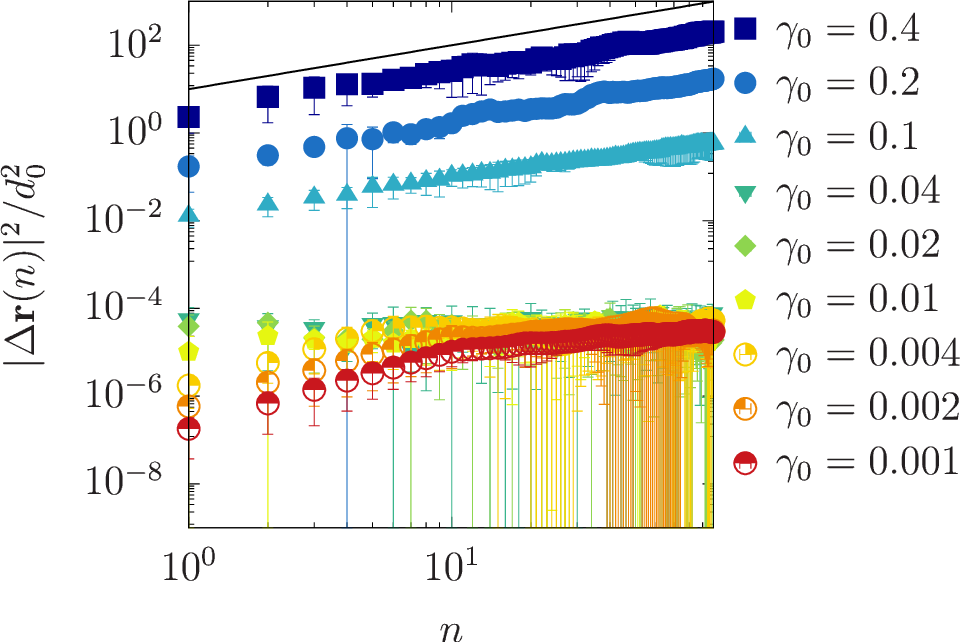}
  \caption{Plots of the mean square displacement $|\Delta {\mathbf r}(n)|^2$ versus $n$ for various values of $\gamma_0$ with $\mu=1.0$. The solid line represents $|\Delta {\mathbf r}(n)|^2 \sim n$.}
\label{msd}
\end{figure}

In Fig. \ref{msd}, we plot the mean square displacements for various values of $\gamma_0$ with $\mu=1.0$ and $n_0 = 100$, defined by
\begin{equation}
  |\Delta {\mathbf r}(n)|^2 = \sum_i |\boldsymbol r_i((n + n_0)T) - \boldsymbol r_i(n_0 T)|^2/N.
\end{equation}
Here, the position $\boldsymbol r_i(n_0 T)$ after $n_0$ cycles is the reference state.
For $\gamma_0 = 0.4, 0.2,$ and $0.1$, $|\Delta {\mathbf r}(n)|^2$ is proportional to $n$, whereas $|\Delta {\mathbf r}(n)|^2$ reaches a small saturated value for $\gamma_0 \le 0.04$.
These results are consistent with the behavior shown in Fig. \ref{dr}, where the system is irreversible for $\gamma_0\ge 0.1$, and reversible for $\gamma_0 \le 0.04$.

From Figs. \ref{dr} and \ref{msd}, we define the reversible phase where the displacement $dr(n)/\gamma_0$ in the last cycle is lower than $0.01 d_0$, and the diffusion coefficient $D$ is lower than $1.0 \times 10^{-5} d_0^2/t_0$.
Here, we estimate $D$ from the mean square displacement $|\Delta {\mathbf r}(n)|^2$ for $n_{\rm I}\le n \le n_{\rm F}$ as $D = (|\Delta {\mathbf r}(n_{\rm F})|^2 - |\Delta {\mathbf r}(n_{\rm I})|^2)/\left \{ 4(n_{\rm F}-n_{\rm I}) \right \}$, where $n_{\rm I} = 10$ and $n_{\rm F} = 100$.
It should be noted that $|\Delta {\mathbf r}(n)|^2$ in the reversible phase is subdiffusive, where $D$ decreases as $n_{\rm F}$ increases.
We confirmed that the systems with $\mu=0.01, 0.1, 0.2, 0.3, 0.5,$ and $1.0$ are in the reversible phase for $\gamma_0 \le 0.04$. \footnote{We have also checked that particles in the reversible phase exhibit almost the same trajectories for ten cycles in all samples.}

\section{Particle trajectories in the reversible phase}
\label{Loop}

\begin{figure}[htbp]
  \begin{center}
    \begin{tabular}{c}
      \begin{minipage}{0.5\hsize}
      \begin{center}
        \includegraphics[width=1.0\linewidth]{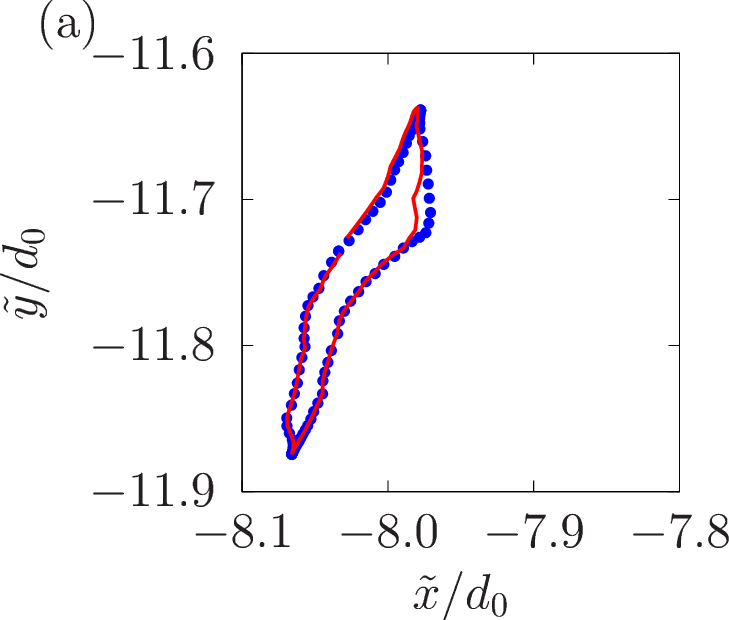}
      \end{center}
      \end{minipage}
      \begin{minipage}{0.50\hsize}
      \begin{center}
        \includegraphics[width=1.0\linewidth]{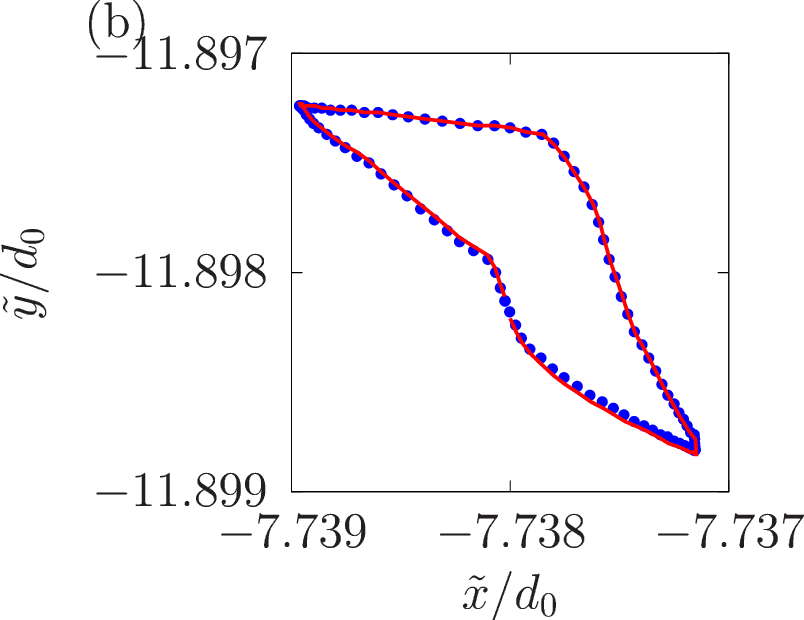}
      \end{center}
      \end{minipage}
    \end{tabular}
\caption{
  Non-affine particle trajectories for $\gamma_0 = 0.04$ (a) and $0.004$ (b) with $\mu=0.1$. 
    The circles represent the trajectory in the last cycle. The line represents the trajectory in the second to the last cycle.
}
\label{trjna_mu0.1}
  \end{center}
\end{figure}

In the reversible phase, non-affine particle trajectories should form closed loops, where the non-affine trajectory for particle $i$ is defined as
\begin{equation}
  \tilde {\boldsymbol r}_i(t) = {\boldsymbol r}_i(t) - \gamma(t) y_i(t) \boldsymbol e_x.
\end{equation}
We plot $\tilde {\boldsymbol r}_i(t)$ for the last two cycles for $\gamma=0.04$ and $0.004$ with $\mu=0.1$ in Fig. \ref{trjna_mu0.1}.
The particle returns to its original position after each cycle, and the trajectory forms a loop; however,the trajectories of two cycles for $\gamma_0=0.04$ deviate in some parts owing to the inertia of the particles.

\begin{figure}[htbp]
  \begin{center}
    \begin{tabular}{c}
      \begin{minipage}{0.5\hsize}
      \begin{center}
        \includegraphics[width=1.0\linewidth]{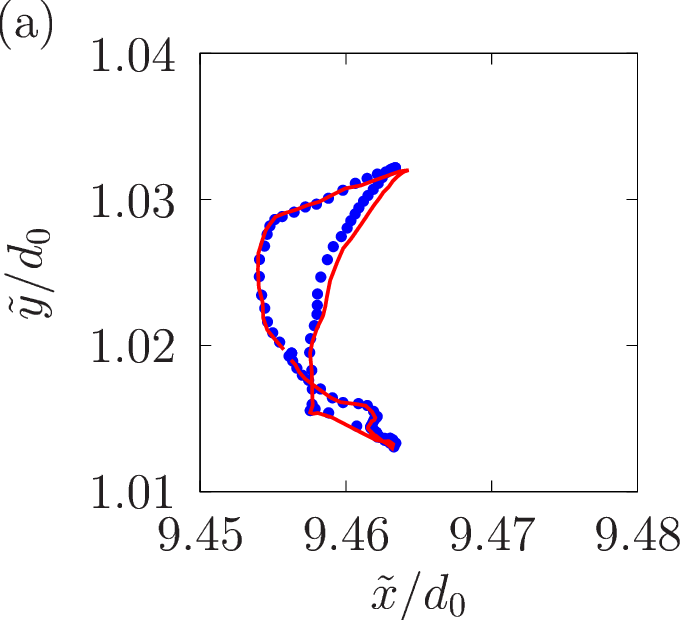}
      \end{center}
      \end{minipage}
      \begin{minipage}{0.50\hsize}
      \begin{center}
        \includegraphics[width=1.0\linewidth]{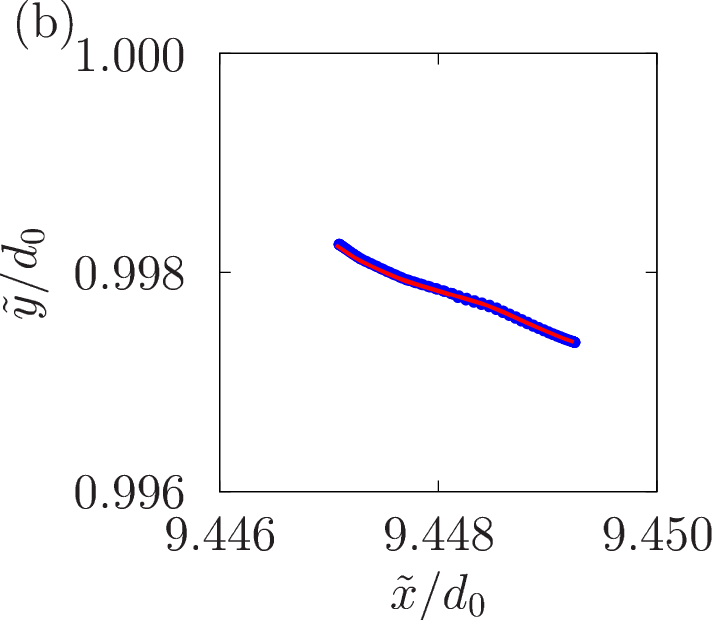}
      \end{center}
      \end{minipage}
    \end{tabular}
\caption{
  Non-affine particle trajectories for $\gamma_0 = 0.04$ (a) and $0.004$ (b) with $\mu=0.5$.
    The circles represent the trajectory in the last cycle. The line represents the trajectory in the second to the last cycle.
}
\label{trjna_mu0.5}
  \end{center}
\end{figure}

\begin{figure}[htbp]
  \begin{center}
    \begin{tabular}{c}
      \begin{minipage}{0.5\hsize}
      \begin{center}
        \includegraphics[width=1.0\linewidth]{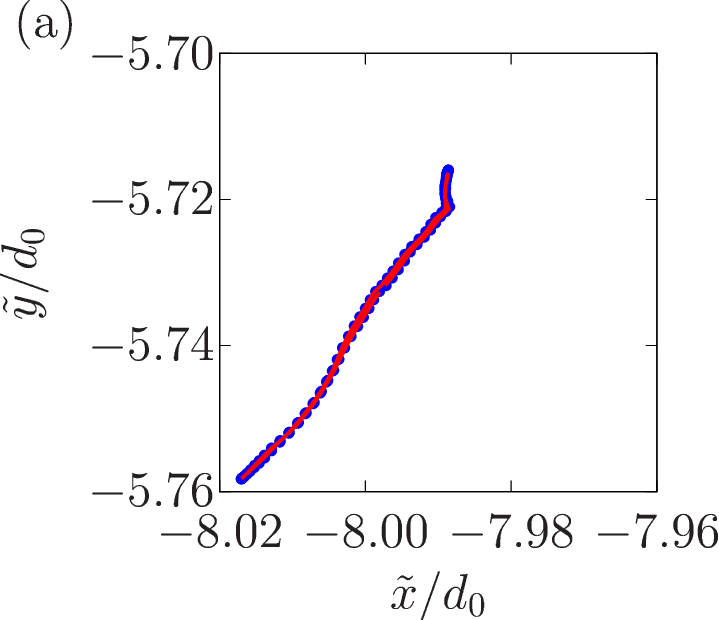}
      \end{center}
      \end{minipage}
      \begin{minipage}{0.50\hsize}
      \begin{center}
        \includegraphics[width=1.0\linewidth]{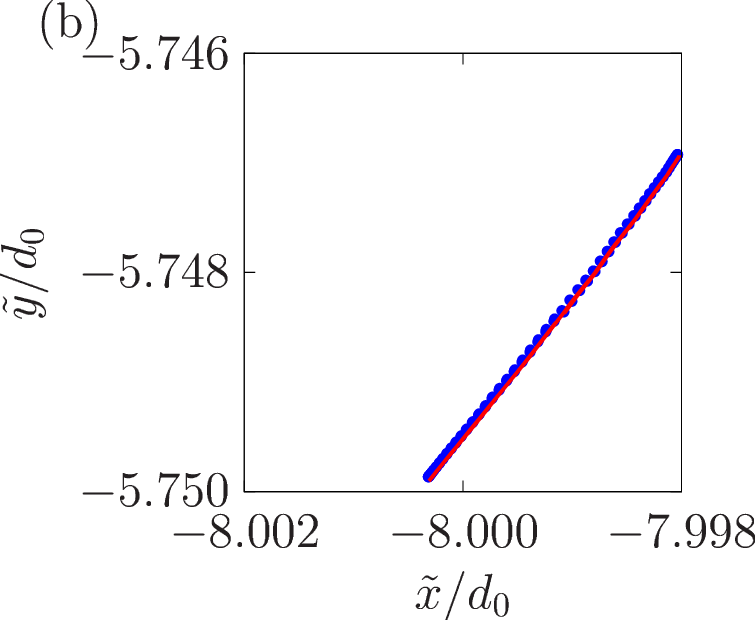}
      \end{center}
      \end{minipage}
    \end{tabular}
\caption{
  Non-affine particle trajectories for $\gamma_0 = 0.04$ (a) and $0.004$ (b) with $\mu=1.0$.
    The circles represent the trajectory in the last cycle. The line represents the trajectory in the second to the last cycle.
}
\label{trjna_mu1.0}
  \end{center}
\end{figure}

In Figs. \ref{trjna_mu0.5} and \ref{trjna_mu1.0}, the non-affine trajectories for $\mu=0.5$ and $1.0$ are plotted.
For $\mu=0.5$, the trajectory forms a loop with a finite area for $\gamma_0 = 0.04$, as shown in Fig. \ref{trjna_mu0.5}(a), while the trajectory for $\gamma_0 = 0.004$ in Fig. \ref{trjna_mu0.5}(b) becomes a line (or a loop with zero area).
Such a line trajectory is not observed in jammed frictionless particles \cite{Lundberg,Schreck,Keim13,Keim14,Regev13,Regev15,Priezjev,Lavrentovich,Nagasawa,Das}.
For $\mu=1.0$, the trajectories for $\gamma_0 = 0.04$ and $0.004$ formed lines, as shown in Fig. \ref{trjna_mu1.0}.
For $\gamma_0 = 0.04$, the line trajectory is bent (Fig. \ref{trjna_mu1.0}(a)), while it is almost straight for $\gamma_0 = 0.004$ (Fig. \ref{trjna_mu1.0}(b)).

\begin{figure}[htbp]
  \includegraphics[width=0.5\textwidth]{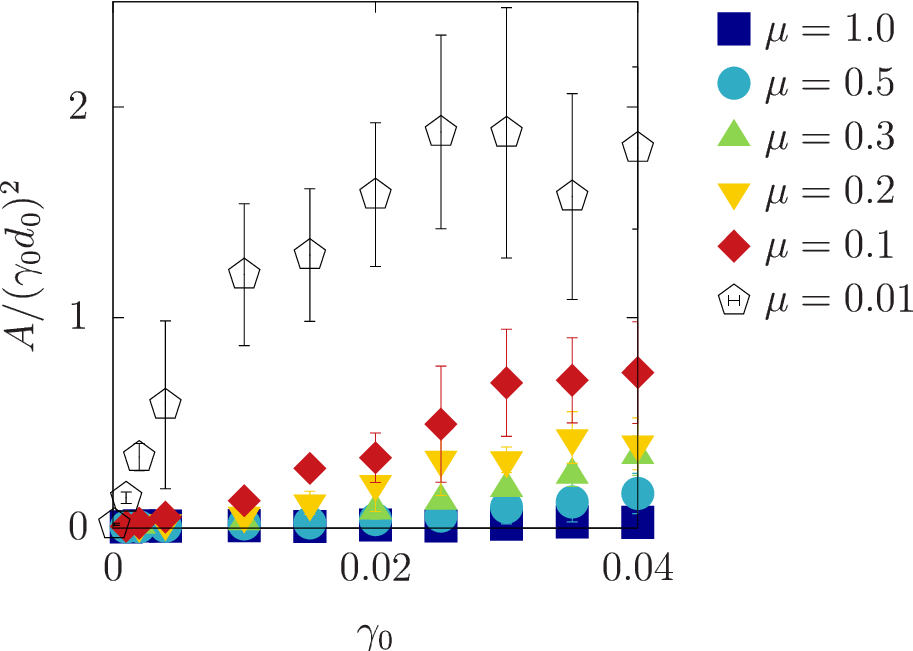}
\caption{
Plots of the normalized area $A/(\gamma_0 d_0)^2$ of loop trajectories versus $\gamma_0$ for various values of $\mu$.
}
\label{A}
\end{figure}

The geometry of the trajectory is characterized by
\begin{equation}
  A_i = \oint_C \tilde x_i d \tilde y_i = \int_0^{2 \pi/\omega} \tilde x_i(t) \frac{d \tilde y_i(t)}{dt} dt,
\end{equation}
where $C$ represents the trajectory of a cycle.
$|A_i|$ coincides with the area covered by the trajectory of particle $i$, if there is no intersection for the trajectory.
We introduce the average area $A$ as
\begin{equation}
  A = \sum_i |A_i| / N.
\end{equation}
If the characteristic length of the trajectory is scaled by $\gamma_0 d_0$, as in frictionless particles \cite{Otsuki21}, $A$ is proportional to $(\gamma_0 d_0)^2$.
Therefore, we plot the normalized average area $A/(\gamma_0 d_0)^2$ against $\gamma_0$ for various values of $\mu$ in Fig. \ref{A}.
When $\gamma_0$ is sufficiently small, $A/(\gamma_0 d_0)^2$ is almost zero, corresponding to the line trajectories for $\mu\ge 0.1$. Furthermore, 
$A/(\gamma_0 d_0)^2$ increases with the strain amplitude $\gamma_0$ above a critical value, which is dependent on $\mu$.

\section{Storage and loss moduli}
\label{Response}

\begin{figure}[htbp]
  \includegraphics[width=0.5\textwidth]{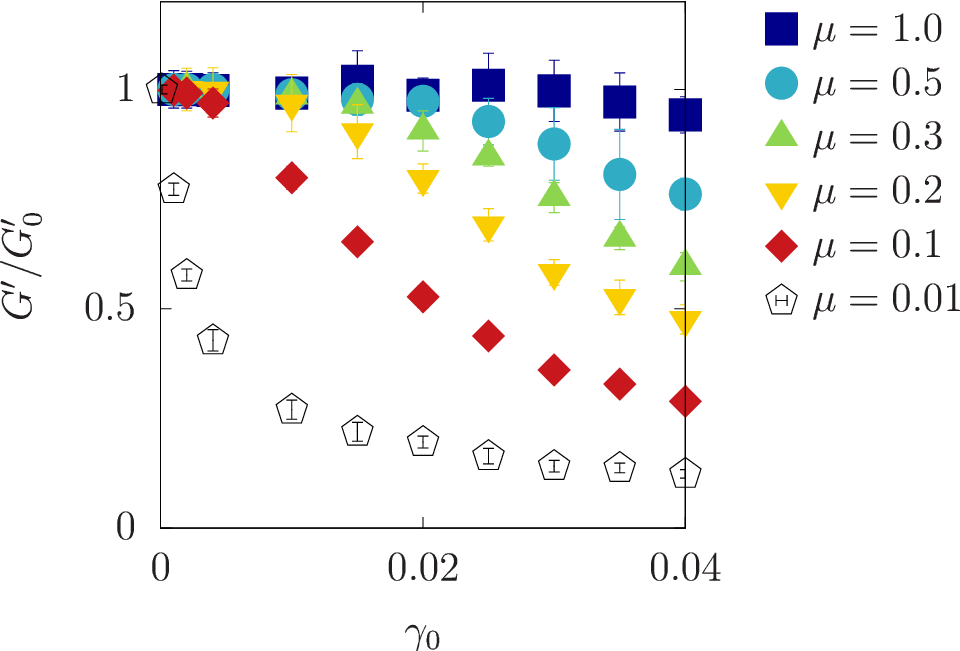}
\caption{
  Plots of the normalized storage modulus $G'/G'_0$ versus $\gamma_0$ for various $\mu$ values.
}
\label{Gp}
\end{figure}

\begin{figure}[htbp]
  \includegraphics[width=0.4\textwidth]{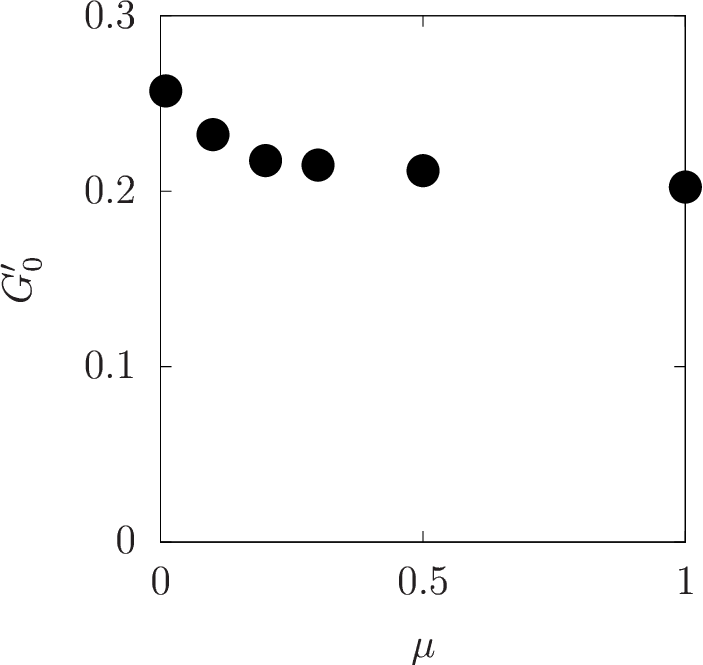}
\caption{
  Plots of the storage modulus $G'_0$ in the linear response regime versus $\mu$.
}
\label{G0}
\end{figure}

In Fig. \ref{Gp}, we plot the scaled storage modulus $G'/G'_0$ in the reversible phase against the strain amplitude $\gamma_0$, where $G'_0$ is defined as the storage modulus $G'_0 = \lim_{\gamma_0 \to 0} G'$ in the linear response regime. 
Here, we estimate $G'_0$ by $G'$ at $\gamma_0 = 1.0 \times 10^{-4}$.
As can be seen in Fig. \ref{Gp},
$G'/G'_0$ decreases with $\gamma_0$ above the critical strain amplitude, depending on $\mu$ for $\mu\ge 0.1$.
The decay of $G'$ with $\gamma_0$ in the reversible phase is regarded as reversible softening, which is also observed in frictionless particles \cite{Otsuki21}.
A similar decay of $G'$ has been reported in an experiment on photoelastic disks, but it is not clear whether it occurs in the reversible phase \cite{Coulais}.
The critical strain amplitude for the decay of $G'$ increases with $\mu$, and the softening decreases as $\mu$ increases. 
In Fig. \ref{G0}, we plot $G'_0$ as a function of $\mu$, where $G'_0$ slowly decreases as $\mu$ increases.

\begin{figure}[htbp]
  \includegraphics[width=0.5\textwidth]{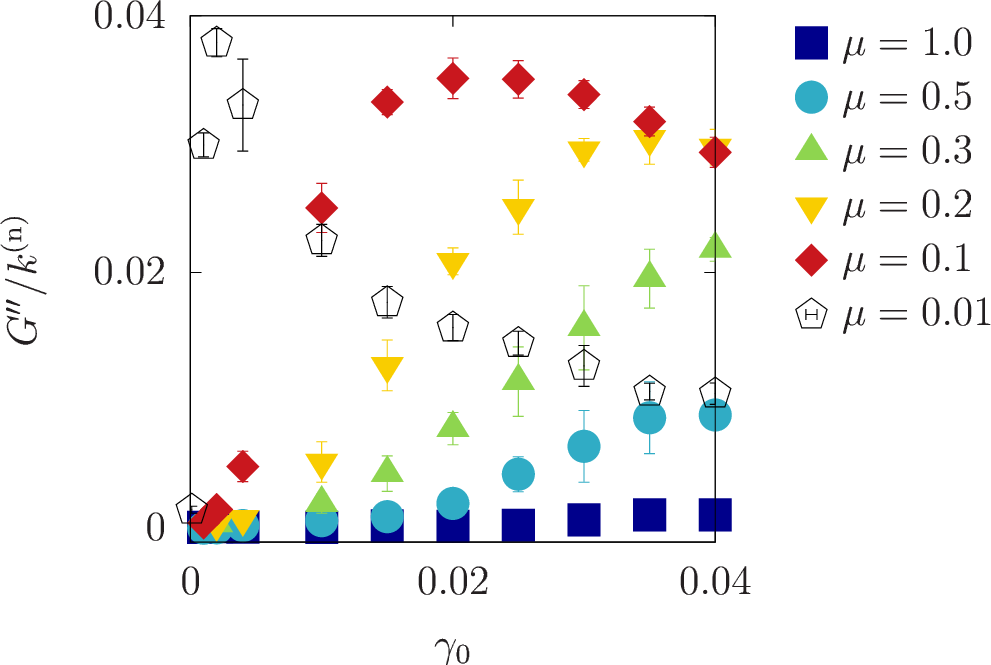}
\caption{
 Plots of the loss modulus $G''$ versus $\gamma_0$ for various values of $\mu$.
}
  \label{Gpp}
\end{figure}

Figure \ref{Gpp} shows the loss modulus $G''$ against $\gamma_0$ for various $\mu$ values.
$G''$ is approximately zero for small $\gamma_0$, while $G''$ increases with $\gamma_0$ above a threshold strain amplitude except for $\mu=0.01$.

\begin{figure}[htbp]
  \includegraphics[width=0.5\textwidth]{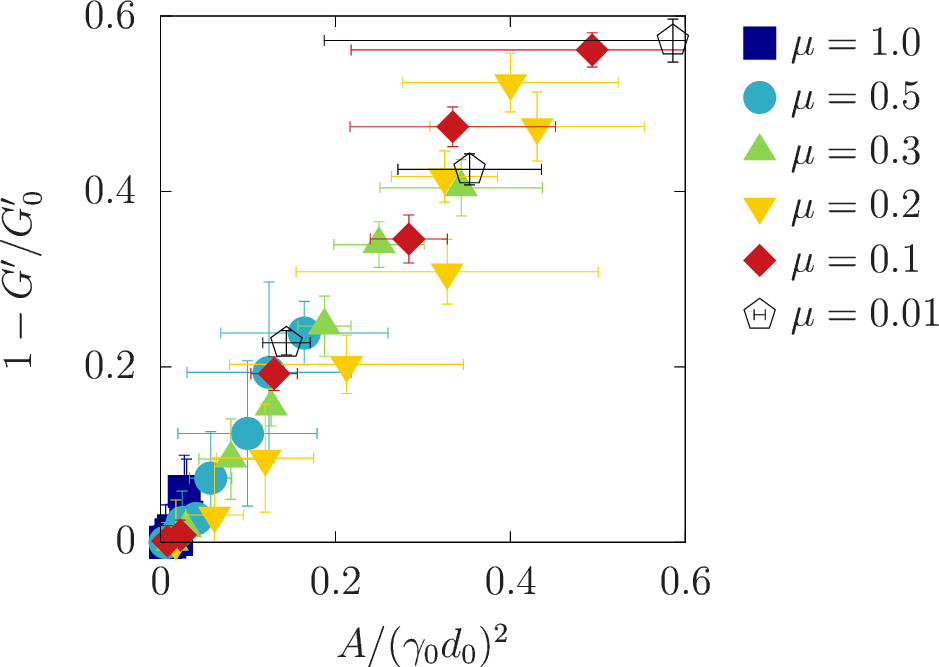}
\caption{
  Plots of $1-G'/G'_0$ versus $A/(\gamma_0 d_0)^2$ for various $\gamma_0$ and $\mu$ values.
}
  \label{Gp_A}
\end{figure}

In Ref. \cite{Otsuki21}, the reversible softening in frictionless systems can be characterized by the loop trajectories of particles.
Even in frictional granular materials, the average area $A$ characterizing the loop trajectories seems to be related to $G'$.
To check the validity of this conjecture, we plotted $1-G'/G'_0$ against $A/(\gamma_0 d_0)^2$ in Fig. \ref{Gp_A} for $A/(\gamma_0 d_0)^2<0.6$.
It is remarkable that $1-G'/G'_0$ satisfies a scaling law in which $1-G'/G_0$ is a linear function of $A/(\gamma_0 d_0)^2$, and is independent of $\mu$.
Note that the data for $A/(\gamma_0 d_0)^2>0.6$ with $\mu=0.1$ and $0.01$ deviate from the linear behavior.

\begin{figure}[htbp]
  \includegraphics[width=0.5\textwidth]{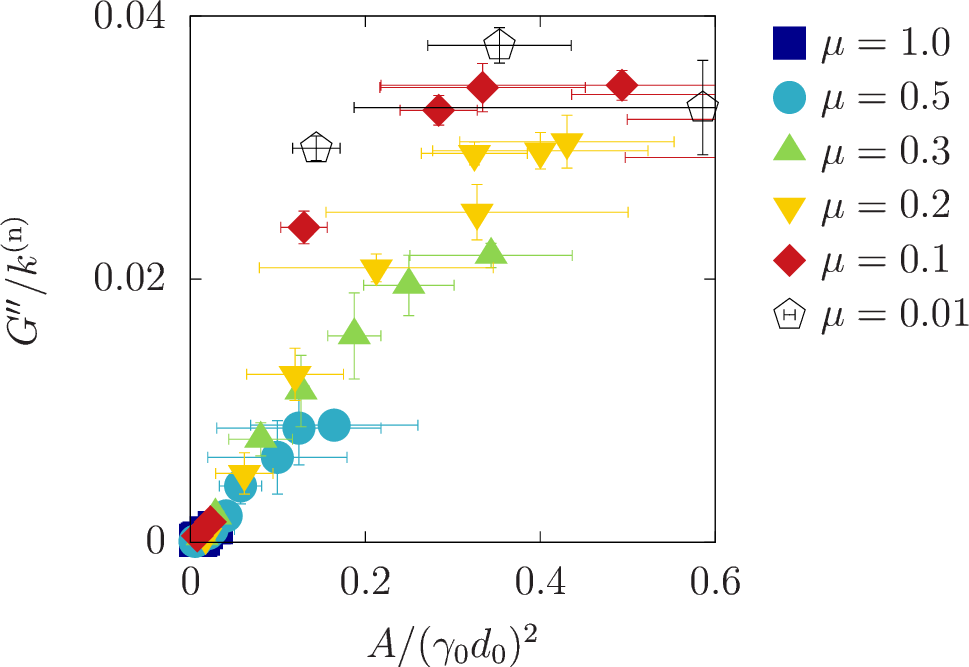}
\caption{
  Plots of the loss modulus $G''$ versus $A/(\gamma_0 d_0)^2$ for various values of $ \gamma_0$ and $ \mu$.
}
  \label{Gpp_A}
\end{figure}

The loss modulus $G''$ is also expected to be characterized by the loop trajectories of the particles even in the frictional system.
The connection between the loss modulus and loop trajectories is suggested in suspension experiments \cite{Keim13,Keim14}.
To clarify this connection, we plotted $G''$ against $A/(\gamma_0 d_0)^2$ in Fig. \ref{Gpp_A} for various $\mu$ with $A/(\gamma_0d_0)^2\le 0.6$.
It is remarkable that $G''$ satisfies a scaling law in which $G''$ is proportional to $A/(\gamma_0d_0)^2$, except for $ \mu=0.1$ and $0.01$.

\section{Conclusion and discussion}
\label{Discussion}

We numerically studied the relationship between the trajectories of the frictional granular materials and the shear modulus under oscillatory shear in the reversible phase.
The geometry of the particle trajectories depends on the friction coefficient $\mu$, where the normalized area $A/(\gamma_0 d_0)^2$ increases as $\gamma_0$ increases and $\mu$ decreases.
The storage modulus $G'$ exhibits reversible softening.
The loss modulus $G''$ remains finite for a large $ \gamma_0$ and small $\mu$.
We found the existence of the scaling laws of $G'$ and $G''$, at least for not too large $A/(\gamma_0d_0)^2$ and not too small $\mu$.

In this study, we investigated the shear modulus of frictional granular materials for $\phi=0.87$.
In future studies, the findings on the shear modulus and loop trajectories will have to be confirmed in the vicinity of the jamming point.

In Ref. \cite{Otsuki21}, the reversible softening and residual loss modulus of frictionless particles are theoretically related to the Fourier components of the loop trajectories.
We numerically connected $G'$ and $G''$ with the area of the loop trajectories, but the theoretical basis has not been confirmed.
Therefore, an extension of the theory in Ref. \cite{Otsuki21} to frictional particles will be our future work.

\begin{acknowledgements}
The authors thank K. Saitoh and D. Ishima for fruitful discussions.
This work was supported by JSPS KAKENHI Grant Numbers JP16H04025, JP19K03670, and JP21H01006, and ISHIZUE 2020 of the Kyoto University Research Development Program.
\end{acknowledgements}



\end{document}